\documentclass[prl,twocolumn,superscriptaddress]{revtex4-2}
\usepackage{amsmath, amstext, amssymb, amsfonts, amsxtra}
\usepackage[usenames]{color}
\usepackage{grffile}
\usepackage{bm}
\usepackage{xcolor}
\usepackage{hyperref}

\newcommand{\zjuone}{State Key Laboratory of Extreme Photonics and Instrumentation, Key Laboratory of Advanced Micro/Nano Electronic Devices and Smart Systems of Zhejiang, Zhejiang University, Hangzhou 310027, China}
\newcommand{\zjutwo}{International Joint Innovation Center, The Electromagnetics Academy at Zhejiang University, Zhejiang University, Haining 314400, China}
\newcommand{\zjuthree}{Jinhua Institute of Zhejiang University, Zhejiang University, Jinhua 321099, China}
\newcommand{\nusone}{Department of Electrical and Computer Engineering, National University of Singapore, Singapore 117583, Singapore}
\begin{document}

\title{Scattering symmetry of diffusive systems}

\author{Dong Wang}
\thanks{These authors contributed equally to this work.}
\affiliation{\zjuone}
\affiliation{\zjutwo}
\affiliation{\zjuthree}

\author{Pei-Chao Cao}
\thanks{These authors contributed equally to this work.}
\affiliation{\zjuone}
\affiliation{\zjutwo}
\affiliation{\zjuthree}

\author{Yanxiang Wang}
\affiliation{\zjuone}
\affiliation{\zjutwo}
\affiliation{\zjuthree}

\author{Minghong Qi}
\affiliation{\zjuone}
\affiliation{\zjutwo}
\affiliation{\zjuthree}

\author{Ran Ju}
\affiliation{\zjuone}
\affiliation{\zjutwo}
\affiliation{\zjuthree}

\author{Hongsheng Chen}
\email{hansomchen@zju.edu.cn}
\affiliation{\zjuone}
\affiliation{\zjutwo}
\affiliation{\zjuthree}

\author{Cheng-Wei Qiu}
\email{chengwei.qiu@nus.edu.sg}
\affiliation{\nusone}

\author{Ying Li}
\email{eleying@zju.edu.cn}
\affiliation{\zjuone}
\affiliation{\zjutwo}
\affiliation{\zjuthree}




\begin{abstract}
Significant progress in manipulating heat diffusion has been achieved with the advent of non-Hermitian physics and topology. However, previous studies on diffusive systems have primarily concentrated on isolated cases, where fields decay exponentially over time. In practical scenarios, systems inevitably interact with external environments, making it essential to study their responses to external heat signals. This, in turn, relies on analyzing the scattering behavior of these signals. In our work, we experimentally realize thermal scattering in a diffusive anti-parity-time (APT) system. We define key parameters of the temperature field—amplitude, phase, and chirality—and reveal that the scattering symmetry of the APT diffusive system only arises when temperature signals with different chiralities interact. Such mechanism is induced by the unique dispersion properties in diffusive systems, where positive and negative frequencies are inequivalent, corresponding to different chiralities. This also explains the difficulty of observing APT scattering symmetry in wave systems. Our findings highlight the pivotal role of scattering channels in the symmetry and phase transitions of non-Hermitian systems and propose novel approaches for analyzing and controlling strongly dissipative phenomena.
\end{abstract}


\maketitle
Manipulating heat assumes a critical role optimizing thermal utilization ~\cite{Alva2018,Yang2024}, advanced thermal management, ~\cite{Moore2014,Hu2020,Liang2022,Zhou2023}, and infrared signal processing ~\cite{Li2018,Peng2020}. Such challenges can be categorized into three distinct paradigms. The first involves the control of heat transfer at steady state or zero frequency ($\omega=0$). The objective of such studies is to achieve the desired spatial distributions of temperature or heat flux, with a typical example being thermal metamaterials ~\cite{Li2021,Yang2021,Yang2024} aiming to achieve extraordinary properties beyond natural materials, such as thermal cloaking ~\cite{Xu2014,Han2014,Peralta2020} and transparency ~\cite{Xu2019,Woo2022}, thermal concentration ~\cite{Li2016,Jin2023} and ultra-conductivity ~\cite{Guo2022}, thermal dome ~\cite{Zhou2024}, thermal nonreciprocity ~\cite{Lei2024} and thermal coherent perfect absorption ~\cite{Li2022}. The theory behind is the transformation theory ~\cite{Hu2018,Su2021}. Recently, with the bloom of machine learning, the AI-assisted design of thermal metamaterials has also been proposed and realized ~\cite{Ji2022,Wang2023,Jin2024,Xing2024}.

The second paradigm investigates the comprehensive spatiotemporal evolution within isolated thermal fields to reveal novel transient heat transfer behaviors. Given the dissipative nature of the system under study, the frequency assumes a complex or purely imaginary number (Im$(\omega)\neq 0$). Utilizing effective Hamiltonian approaches in heat transfer ~\cite{Li2019,Li2025}, this framework has unearthed an abundance of exotic  phenomena, including the heat-locking effect ~\cite{Li2019}, phase oscillation ~\cite{Cao2024}, topological transition ~\cite{Xu2021,Xu2022}, and localized heat diffusion ~\cite{Cao2021,Qi2022}. Due to inherent dissipation, the effective Hamiltonian of diffusive systems is naturally non-Hermitian, which provides the possibility to explore novel traits with the methodology used in non-Hermitian physics ~\cite{Miri2019,ElGanainy2018,Oezdemir2019,Gupta2020,Ashida2020,Bergholtz2021,Huang2024,Mao2023,2023,Wang2024a}. For instance, the aforementioned heat-locking effect is protected by the anti-parity-time (APT) symmetry ~\cite{Li2019}. Regarding parity-time (PT) symmetry, recent observations in diffusive systems have demonstrated its emergence through the realization of convection-induced real coupling, an effect that contributes to the suppression of thermal phase oscillations. ~\cite{Cao2024}. Furthermore, the non-Hermitian thermal topology associated with Weyl exceptional rings (WERs) has also been elucidated. ~\cite{Xu2022a}.

While the aforementioned studies have unveiled unique non-Hermitian characteristics, their scope is confined to isolated systems and analyzed primarily from the perspective of the Hamiltonian formalism. In realistic scenarios, however, systems invariably interact with their environments, where scattering processes assume a pivotal role in characterizing these interactions. The scattering of external signals by the systems can encapsulate richer non-Hermitian physics and interesting phenomena. For instance, the phenomena of coherent perfect absorption of light (or anti-laser action) and unidirectional invisibility ~\cite{Lin2011} have been demonstrated to be characterized by the PT symmetry of the scattering matrix and predicted through the phase transition of its eigenvalues ~\cite{Chong2011}. Therefore, it is crucial to investigate a third paradigm of heat manipulation, specifically one targeting time-harmonic temperature fields carrying heat signals (Re$(\omega)\neq0$), to evaluate scattering behaviors of heat. In particular, this investigation can reveal the symmetry of scattering, which, as we will elaborate later, is fundamentally different from the symmetry of the scatterer itself. Yet, due to the lack of comprehensive analytical frameworks and effective experimental platforms, studies in this area remain scarce and largely theoretical ~\cite{Li2021a,Ju2024}. 

Here, we established a systematic theoretical framework for thermal scattering phenomena, revealing that scattering symmetry is fundamentally tied to time-reversal anti-symmetry (anti-T symmetry) of heat signals in diffusive channels. This connection links the two chiralities of thermal fields. Based on this, we identified an APT-symmetric thermal scattering system and obtained its phase diagram. Furthermore, we constructed a full experimental platform that enables the realization of real-frequency resonant thermal signals with tunable chirality. Through this platform, we observed—for the first time—the APT thermal scattering phenomenon, which exhibits a phase transition leading to one-sided heat suppression. Our work underscores the transformative potential of scattering-based approaches for manipulating heat, opening avenues for future advancements in thermal signal control based on scattering symmetries.

\begin{figure}[tbp]
\centering
\includegraphics[width=\columnwidth]{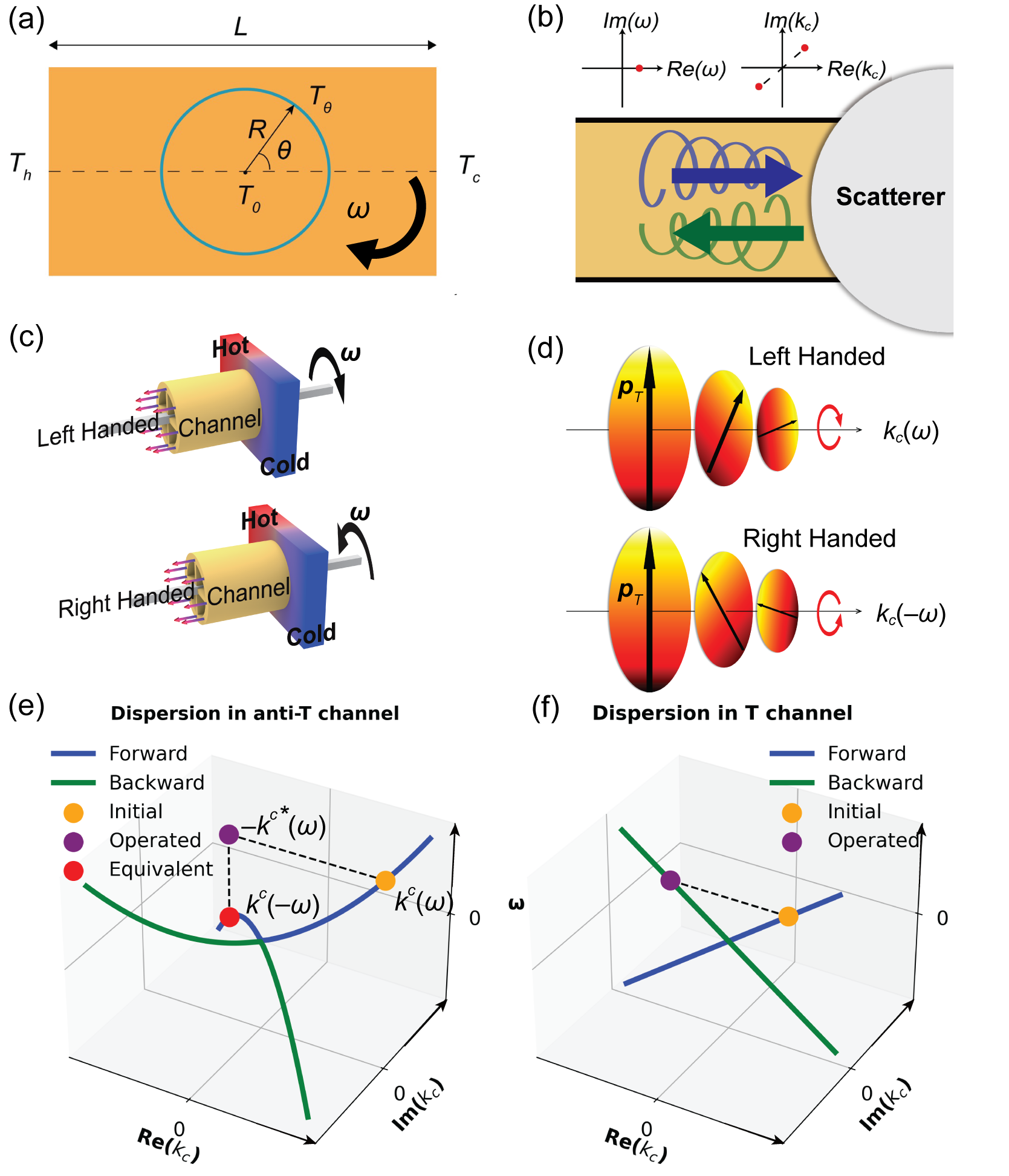}
\caption{Thermal scattering and symmetry.
(a) Generation of time-harmonic thermal signals with frequency $\omega$. Distinct rotation directions correspond to different chiralities. The resulting signal is expressed as $T(x,z,t)=Ae^{ik_{c}z}e^{ikx\pm i\omega t}$ (b) Spatial separation of forward- and backward-propagating thermal signals. (c) Experimental realization of chiral heat sources. (d) Schematic illustration of LH and RH signal propagation. (e) Dispersion relation in an anti-T symmetric scattering channel. The initial state (yellow sphere) transforms under time reversal to the purple sphere, physically equivalent to the red sphere. Blue and green curves depict forward and backward waves, respectively. (f) Dispersion relation in a T symmetric channel, characterized by ${k_c}^2 = {\omega}^2/c^2$, where $c$ is the wave velocity. The time-reversed state (purple sphere) corresponds to a physical state.}

\label{fig:figure1}
\end{figure}

To generate time-harmonic thermal signals, we project a linear temperature profile onto a circular geometry and impose a rotational modulation. Consider a rectangular thermally conductive plate, as depicted in Fig.~\ref{fig:figure1}(a), with two opposing sides maintained at fixed temperatures $T_h$ and $T_c$, where $T_h > T_c$. The temperature distribution across the plate is then described by $T(R,\theta) = T_0 + \nabla T R \cos\theta$, where $T_0 = (T_h + T_c)/2$ denotes the mean temperature and $\nabla T = (T_h - T_c)/L$ is the temperature gradient along the plate of length $L$. By applying a constant angular velocity $\omega$ to the plate, the temperature at a location $(R,\phi)$ in the laboratory frame transforms to that at $(R, \theta + \omega t)$ in the rotating frame, yielding $T(R,\theta,t) = T_0 + \nabla T R \cos(\theta + \omega t)$. Expressing the spatial coordinate on the circle as $x$ and converting the trigonometric expression to its Euler representation, we obtain a heat source with a real frequency $\omega$ and amplitude $A = \nabla T R$: $T(x,t) = T_0 + A e^{i k_x x + i \omega t}$, where $k_x = 1/R$ is the wavevector.

Introducing an oscillating heat source into a diffusive channel enables the excitation of time-harmonic heat signals. Within such channels, the temperature field is governed by:
\begin{equation}
	\partial_t T=a^2 \nabla^2 T
	\label{eq:1}
\end{equation}
\noindent where $a^2 = D_c$ represents the thermal diffusivity within the channels. Assuming the heat signal exhibits spatially separated forward and backward propagating components (considering $z$ as the direction of propagation), we adopt a time-harmonic ansatz for the excited heat signal: $T_c(x,z,t)=F_c(z)e^{ik_xx-i\omega t}$, with the spatially dependent component given by $F_c(z)=B_{c+}e^{ik_c z}+B_{c-}e^{-ik_c z}$. Substituting this expression for $T_c(x,z,t)$ into the heat diffusion governing equation (Eq.~\ref{eq:1}), the propagation wavenumber $k_c$ is readily determined:
\begin{equation}
	{k_c}^2=-{k_x}^2+i\omega/D_c
	\label{eq:2}
\end{equation}

\noindent In conventional heat transfer, the time-dependent term $e^{-i\omega t}$ is purely dissipative and typically expressed as $e^{-\alpha t}(\alpha>0, \alpha\in\mathbb{R})$. The implies the frequency $\omega=-i\alpha$ is a purely imaginary number,  precluding the existence of a phase velocity, defined as $v_p=\text{Re}(\omega)/\text{Re}(k_c)$. Absent a propagating phase, the positive and negative propagation wavenumbers $\pm k_c$ in ~Eq.~\ref{eq:2} do not genuinely represent opposing propagation directions, thus hindering the establishment of a pertinent scattering model for heat signals. To facilitate propagation, both the frequency and wavenumber must possess real components, which emphasizes the necessity of constructing a heat source with real frequencies again. A real-value $\omega$ in ~Eq.~\ref{eq:2} dictates a complex propagation wavenumber $k_c$, leading to a decay in the excited heat signals along the propagation path due to its imaginary component, as illustrated in ~Fig.~\ref{fig:figure1}(b). This decay facilitates the separation of forward and backward propagating signals.

The implementation of heat source with real frequencies heralds the prospect of defining chirality in heat signal. Drawing an analogy to the circular polarization of electromagnetic waves, the chirality of heat signals can be defined via the thermal dipole $\boldsymbol{p}_T$ in cross section: $\boldsymbol{p}_T=\iint \boldsymbol{r} \sigma_T d S=i \iint x \sigma_T d S+j \iint y \sigma_T d S$,
\noindent where $\sigma_T$ denotes the temperature configuration within the cross section. By modulating the rotational direction of our engineered heat source, we engendered the Left-Handed (LH) and Right-Handed (RH) signals, as depicted in ~Fig.~\ref{fig:figure1}(c) and~Fig.~\ref{fig:figure1}(d). The LH (RH) signifies that the rotational direction of the thermal dipole and the direction of propagation adhere to the left-hand (right-hand) rule.

Prior to examining the symmetry of scattered heat signals, we must first revisit the symmetry of propagating wave signals (e.g., electromagnetic waves) during propagating. Within a lossless medium, such as a vacuum, the wave field $F$ typically obeys the wave equation $\partial_t^2 T=a^2 \nabla^2 T$, where $a$ is a real number. The second-order time derivative inherently satisfies time-reversal (T) symmetry. Notably, replacing the second-order time derivative with a first-order counterpart $\partial_t^2 F \rightarrow \partial_t F$, yields the diffusive equation ~Eq.~\ref{eq:1}. The diffusive equation, however, exhibits time-reversal anti-symmetry (anti-T) rather than T-symmetry. To elucidate this, we express the Hamiltonian governing the diffusive system ~Eq.~\ref{eq:1} as $H_c=i a^2 \nabla^2$, analogous to the Schr$\ddot{\text{o}}$dinger equation form: $i\partial_t F=H_c F$.  It is evident that the Hamiltonian $H_c$ and time reversal operator satisfy anti-commutation relation $\{H_c,\mathcal{T}\}=0$, thus characterizing this symmetry as anti-T-symmetry. The T-symmetry and anti-T-symmetry in scattering channels manifest in the dispersion relation of scattered signals, as depicted in ~Fig.~\ref{fig:figure1}(e) and ~Fig.~\ref{fig:figure1}(f). Since the wavenumbers in anti-T channels are complex for real $\omega$, time-reversal maps $k_c(\omega)$ to $-k_c^{*}(\omega)$. However, the state ($-k_c^{*}$) does not lie on the dispersion curve at the same frequency, indicating it is not a physical state. Instead, there is an equivalent state, $k_c(-\omega)$, corresponding to $-k_c^{*}(\omega)$, that resides on another dispersion branch with negative frequency.

Incorporating these findings with the construction of heat signals with chiralities, we reveal that the state $k_c(\omega)$ before time-reversal represents left-handed (LH) signals, while the state $k_c(-\omega)$ after time-reversal corresponds to right-handed (RH) signals. Consequently, chiralities are intertwined in anti-T scattering channels. It is essential to note that this linkage is unique to diffusive systems, as in wave systems, positive and negative frequencies are equivalent (Fig. \ref{fig:figure1}(f)).

For general two-port thermal scattering, the temperature field in channels takes the form $T_{ci}=\left[A_i u_i^{\text {in }}(z, \omega)+B_i u_i^{\text {out }}(z, \omega)\right]$, where $u_i^{\text{in}}$ and $u_i^{\text{out}}$ denote the incoming and outgoing modes, respectively, and $A_i$ and $B_i$ are their corresponding amplitudes. The scattering matrix $\boldsymbol{S}(\omega)$ relates $\boldsymbol{B}$ and $\boldsymbol{A}$ through $\boldsymbol{B} = \boldsymbol{S}(\omega) \boldsymbol{A}$, with $\boldsymbol{A} = \left[A_1, A_2\right]^{\mathrm{T}}$ and $\boldsymbol{B} = \left[B_1, B_2\right]^{\mathrm{T}}$. By applying $\mathcal{PT}$ operators on both sides of $T_{ci}$, we eventually derive the form (see Supplementary Materials, the channel's symmetry is used): $\left(\mathcal{PT} T_c\right)_i=(\mathcal{PT} A)_i u_i^{\text {in }}(z, -\omega)+(\mathcal{PT}B)_i u_i^{\text {out }}(z, -\omega)$. For APT symmetric systems, if an eigenstate $T_{ci}$ exists at frequency $\omega$, there must be a corresponding eigenstate $(\mathcal{PT} T_c)_i$ at frequency $-\omega^*$. Since $\omega$ is real, the temperature field undergoing $\mathcal{PT}$ transformation should exhibit another scattering relationship characterized by $\boldsymbol{S}(-\omega)$, where $\mathcal{PT} \boldsymbol{B} = \boldsymbol{S}(-\omega) \mathcal{PT}\boldsymbol{A}$. By consolidating the scattering relation $\boldsymbol{B}=\boldsymbol{S}(\omega)\boldsymbol{A}$, we can derive the symmetry of scattering matrix in APT systems as follows:

\begin{figure}[b]
\centering
\includegraphics[width=\columnwidth]{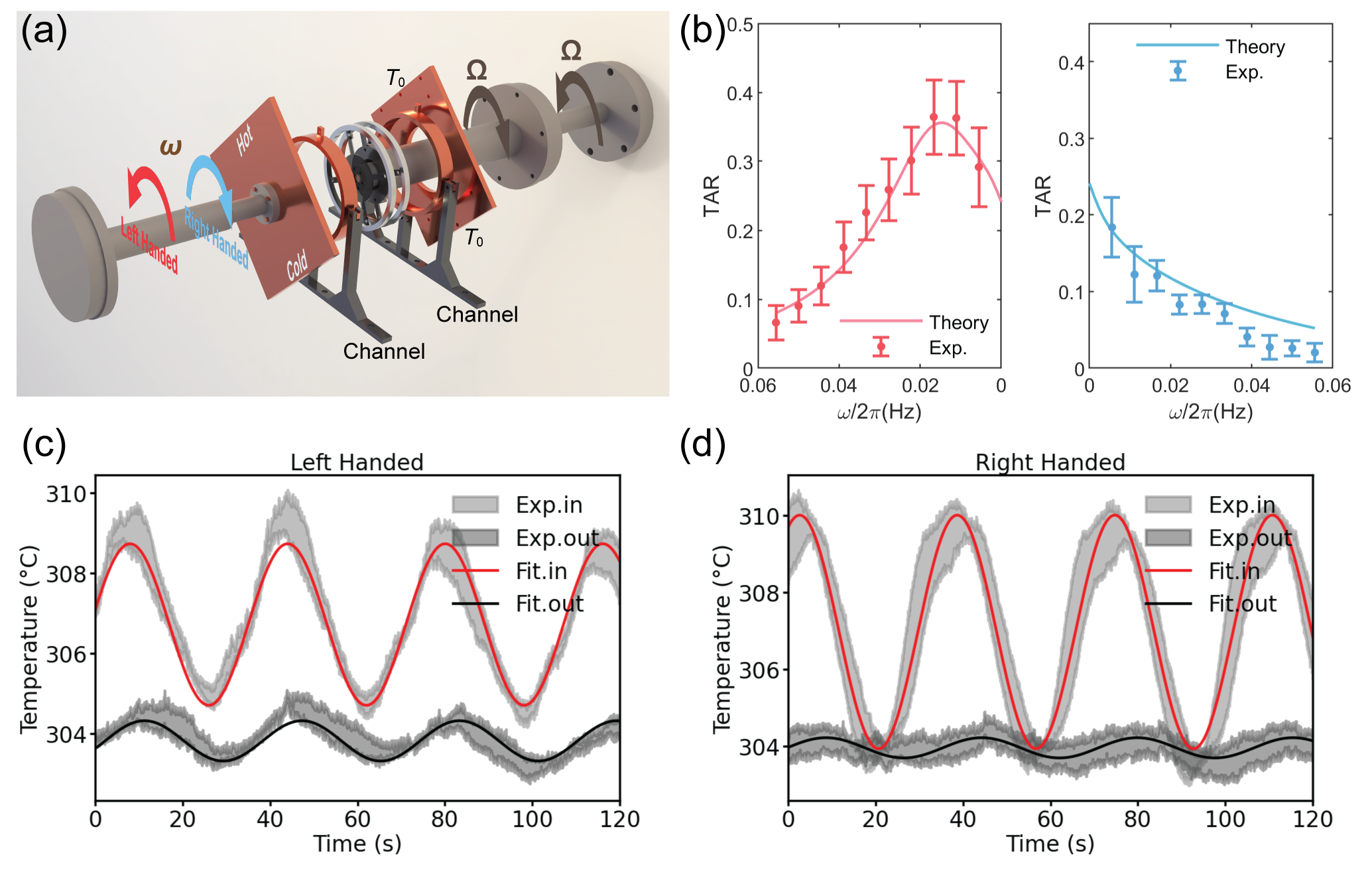}
\caption{Experimental realization of thermal scattering from an anti parity-time symmetric(APT) scatterer with Left Handed(LH) signals and Right Handed(RH) signals.
(a) Experimental setup for achieving thermal scattering. The apparatus comprises three rotators: the leftmost rotator controls the frequency and chirality of the heat source, while the two rightmost rotators control the internal parameter ($\Omega$) of the APT scatterer. The copper plane on the left serves as a heat source with real frequency, and the copper plane on the right serves as a uniform heat source at temperature $T_0$.(b) Temperature Amplitude Ratio (TAR) spectra for LH (left panel) and RH (right panel) heat signals. The experiment was conducted with $\Omega=1.67$ rpm. (c, d) Temperature fields for LH/RH signals. Red curves represent the temperature profiles at the left port, while black lines represent those at the right port. The excitation frequency is $0.028$ Hz.}
\label{fig:figure3} 
\end{figure}

\begin{equation}
	(\mathcal{PT}) \boldsymbol{S}(\omega)(\mathcal{PT})^{-1}=\boldsymbol{S}(-\omega)
	\label{eq:9}
\end{equation}
\noindent where $\boldsymbol{S}(\omega)$ describes the scattering for LH signals and $\boldsymbol{S}(-\omega)$ represents the scattering for RH signals. For clarity, we shall denote them $\boldsymbol{S}_L$ and $\boldsymbol{S}_R$ respectively.

The symmetry of scattering in APT systems is deposited into two matrices with LH and RH chiralities, which also build a bridge linking different chiralities in heat signals. To experimentally observe this scattering symmetry, both LH and RH signals must be excited. ~Fig.~ \ref{fig:figure3}(a) delineates the rotating structures employed in our experiment to realize thermal scattering, comprising two coupled thermal resonators that constitute the APT scatterer (see Supplementary Materials). Ring-shaped channels are utilized to propagate thermal signals with specific frequencies and chiralities. We use the normalized input $\left(1, 0\right)^{\mathrm{T}}$ as an excitation to this two-port scattering model, wherein the ``1'' represents $T_0 + Ae^{ikx \pm i\omega t}$ and ``0'' denotes $T_0$. The temperature fields at the left and right ports of the APT scatterer are depicted in ~Fig.~\ref{fig:figure3}(c) and ~Fig.~\ref{fig:figure3}(d). However, these results do not precisely correlate with the transmission coefficients. In theoretical assessments, scattering channels are presumed to be infinite to ensure a singular scattering event (see SM). In practical scenarios, channel lengths are invariably finite, necessitating the consideration of multiple scatterings. To experimentally characterize scattering properties, the temperature amplitudes for both ports were measured, and their ratio, called the Temperature Amplitude Ratio (TAR), was calculated to represent the scatterer's transmission characteristics. ~Fig.~\ref{fig:figure3}(b) presents the TAR for LH and RH signals, demonstrating remarkable concordance with theoretical predictions. In TAR spectrum, the resonant behavior is observed, which can be understood by the Doppler effect that convective scatterer favors the frequency that is close to its eigenfrequency (resonant phenomenon are common in thermal scattering, see SM).

To  mediate the interaction between these chiralities, we introduce a synthetic Hilbert space ($V_\mathrm{co}$) constructed from the tensor product of chirality space ($V_s$) and port space ($V_P$):$V_\mathrm{co}=V_s\otimes V_P$. A deftly designed permutation of the synthetic space for concurrent input is: $\boldsymbol{A}_\text{co} =\left(A^{L}_{1},A^{R}_{2},A^{R}_{1},A^{L}_{2}\right)^{\mathrm{T}}$. The concurrent output $\boldsymbol{B}_\text{co} = \left(B^{R}_{1},B^{L}_{2},B^{L}_{1},B^{R}_{2}\right)^{\mathrm{T}}$ satisfy the relationship $\boldsymbol{B}_\text{co} = \boldsymbol{S}^4 \boldsymbol{A}_\text{co}$, where: 
\begin{equation}
        \boldsymbol{S}^4=P_{\text {out }}\left(\boldsymbol{S}_{L} \oplus \boldsymbol{S}_{R}\right) P_{\text {in }}^T
	\label{eq:10}
\end{equation}
$P_{in}=\frac{1}{2}\left(I^S\otimes\left(I^P+\sigma_z^P\right)+\sigma_x^S \otimes\left(I^P-\sigma_z^P\right)\right)$ and $P_{\text {out }}=\frac{1}{2}\left(\sigma_x^S \otimes\left(I^P+\sigma_z^P\right)+I^S \otimes\left(I^P-\sigma_z^P\right)\right)$ (see SM). Here $(I,\sigma_x,\sigma_Z)^{S/P}$ denotes the Pauli operators in chiality space (L/R) and port space (1/2) and we have used the relation $\boldsymbol{B}^{L/R}=\boldsymbol{S}_{L/R}\boldsymbol{A}^{L/R} $. Using ~Eq.~\ref{eq:9}, we further elucidate the symmetry of the composite scattering matrix:$(\mathcal{PT}) \boldsymbol{S}^4(\mathcal{PT})^{-1}=\boldsymbol{S}^4$. This manifests the invariance of the joint scattering matrix $\boldsymbol{S}^4$ under $\mathcal{PT}$ operation in APT systems.

\begin{figure}[tbp]
\centering
\includegraphics[width=\columnwidth]{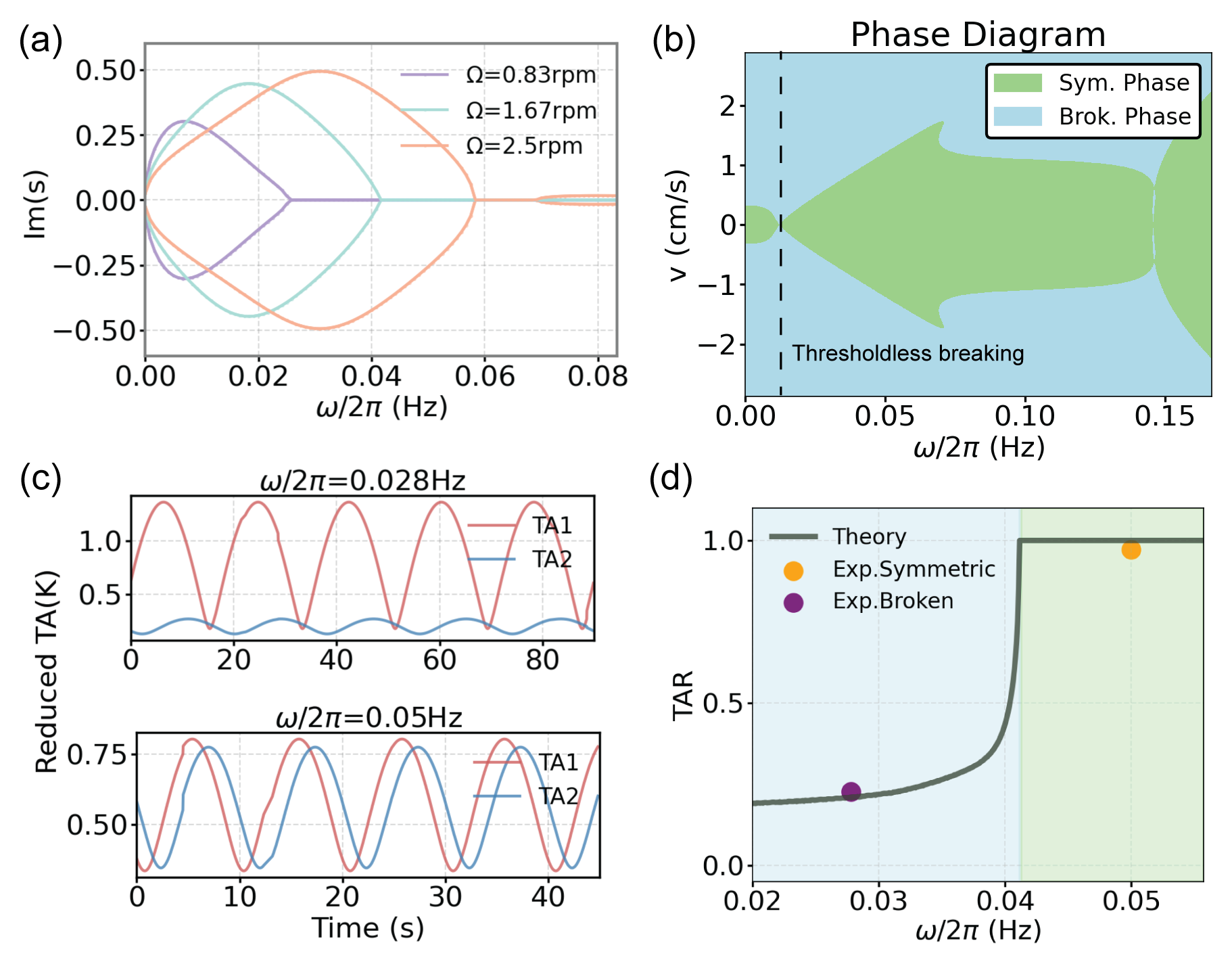} 
\caption{
(a) Imaginary eigenspectra of the compound scattering matrix $\boldsymbol{S}^4$ at varying rotational velocities. (b)Phase diagram of the scattering matrix, wherein the black dashed line demarcates a symmetry-broken phase, independent of internal parameters, for $v = \Omega R$, where $R$ denotes the scatterer's radius. (c) Temporal evolution of the temperature amplitude at both the left (red curve) and right (blue curve) ports. The driving frequency for the upper panel is $0.028$ Hz, corresponding to the symmetry-broken phase, whereas the lower panel features a frequency of $0.056$ Hz, indicative of the symmetric phase. (d) Phase transition dynamics, characterized by the ratio of the mean temperature amplitudes at the left and right ports. Orange and purple data points correlate with the upper and lower panels in (c), respectively.}
\label{fig:figure4}
\end{figure}

To now, we have successfully observed the APT scattering symmetry, $[\mathcal{PT},\boldsymbol{S}^4]=0$, in APT systems characterized by $\{\mathcal{PT},H \}=0$. This relationship corroborates the notion that the symmetry of scattering is different from the symmetry of Hamiltonian, as the former requires both the symmetry of scatterer (Hamiltonian) and scattering channels. In PT systems, the scattering symmetry has been observed in the form of $(\mathcal{PT}) \boldsymbol{S}(\omega)(\mathcal{PT})^{-1}=\boldsymbol{S}^{-1}(\omega^*)$ ~\cite{Chong2011}, the symmetry breaking would lead to the emergence of coherent perfect absorption (CPA). For APT systems, the scattering symmetry lies on $(\mathcal{PT}) \boldsymbol{S}(\omega)(\mathcal{PT})^{-1}=\boldsymbol{S}(-\omega^*)$, which is essentially different because it involves the interplay of different chiralities.

The eigenvalues $s$ of the joint scattering matrix $\boldsymbol{S}^4$ is:
\begin{equation}
	s^2=\operatorname{Re}\left(r_{11} r_{22}{ }^*\right) \pm \sqrt{-\operatorname{Im}\left(r_{11} r_{22}{ }^*\right)^2+|t|^4}
	\label{eq:12}
\end{equation}
\noindent The eigen-spectrum and phase diagram are shown in ~Fig.~\ref{fig:figure4}(a) and ~Fig.~\ref{fig:figure4}(b). The scattering phase transition is governed by both exogenous parameters and endogenous system dynamics. Surprisingly, we observe a peculiar frequency at which scattering symmetry persistently ruptures, irrevocably of the system's intrinsic parameters. From ~Eq.~\ref{eq:12}, we discern that at this singular frequency $|t|^4 =\left[\text{Im}(r_{11}r_{22})\right]^2$ when $\Omega=0$ and $|t|^4 < \left[\text{Im}(r_{11}r_{22})\right]^2$ when $\Omega \neq 0$. The emergence of rotation is responsible for the scattering symmetry-breaking. Given that $\boldsymbol{S}^4$ encapsulates the integral information of this APT symmetric scattering system, this thresholdless symmetry-broken phenomenon could be interpreted as the signal-induced decoupling of the coupled system. This is quite counterintuitive, as most coupled systems possess a critical coupling; only when the coupling exceeds this value can the coupling be decoupled and the symmetry breaking state is achieved. 

To further probe phenomena engendered by scattering symmetry, we consider the input associated with the eigenstates of $\boldsymbol{S}^4$, expressed in vector form: $\Psi^{\mathrm{in}}=\left(s^{L}_{1} e^{-i \omega t}, s^{R}_{2} e^{i \omega t}, s^{R}_{1} e^{i \omega t}, s^{L}_{2} e^{-i \omega t}\right)^{\mathrm{T}}$. The output vector $\Psi^{\mathrm{out}}$ is obtained by the co-scattering matrix $\Psi^{\text {out}}=S^4 \Psi^{\text {in}}$. Since temperature fields for left and right ports comprise respective input and output components, the temperature amplitudes (TA) are $\mathrm{TA}_l=\left|\psi^{\text {in }}_1+\psi^{\text{in}}_3+\psi^{\text{out}}_1+\psi^{\text{out}}_3\right|$ and $\mathrm{TA}_r=\left|\psi^{\text {in}}_2+\psi^{\text{in}}_4+\psi^{\text{out}}_2+\psi^{\text{out}}_4\right|$ respectively. However, direct excitation and experimental realization of this eigenstate prove challenging. An alternative approach is the employment of single-chiral signal superposition. The eigenstate can be decomposed into four discrete single inputs, $|\Psi\rangle=\left(c^{L}_1,c^{R}_2,c^{R}_1,c^{L}_2\right)^{\mathrm{T}}=c^{L}_1 \boldsymbol{e}_1+c^{R}_2 \boldsymbol{e}_2+c^{R}_1 \boldsymbol{e}_3+c^{L}_2 \boldsymbol{e}_4$ where $\boldsymbol{e}_i=\left(\delta_{1i},\delta_{2i},\delta_{3i},\delta_{4i}\right)^{\mathrm{T}}$ denotes the normalized input of a single chiral signal. We measured temperature fields for each individual input, and, invoking the system's linearity, indirectly observed the eigen-input results. In the scattering symmetric phase, $\mathrm{TA}_l$ and $\mathrm{TA}_r$ oscillate at twice the excitation frequency $\omega/2\pi$, with nearly commensurate. Conversely, in the symmetry-broken phase, while both persist to oscillate at the doubled frequency, one amplitude is markedly suppressed. Experimental results are depicted in ~Fig.~\ref{fig:figure4}(c). This phase transition phenomenon is more vividly discernible when evaluating the ratio of their mean values, as illustrated in ~Fig.~\ref{fig:figure4}(d). Prior to the phase transition frequency, the system resides in the scattering-broken phase, wherein one scattering port is significantly subdued. Approaching the phase transition point, this unilateral suppression gradually abates, and upon entering the scattering-symmetric phase, the suppression is entirely eradicated.

Thermal scattering is instrumental in scrutinizing dynamic heat transfer phenomena, particularly in practical contexts where temperature distributions are often transient. As a diagnostic tool, it provides valuable insights into the properties of heat transfer systems, such as identifying internal defects by comparing experimental measurements with theoretical predictions. Notably, while we have presented a specific example of thermal scattering analysis, our theoretical framework is general and can be extended to other heat transfer systems. From an application perspective, one promising direction is investigating the relationship between thermal non-reciprocity and the function of thermal diodes ~\cite{Li2021a,Ju2024}.

\begin{acknowledgments}
The work is supported by the National Key Research and Development Program of China under Grant No. 2023YFB4604100, the National Natural Science Foundation of China (NNSFC)under Grant Nos.12475040 and 52250191, and the Zhejiang Provincial Natural Science Foundation of China under Grant No. LZ24A050002.
\end{acknowledgments}

\bibliography{reference.bib}

\end{document}